\documentclass[
preprint,
superscriptaddress,
showpacs,
longbibliography,
amsmath,
amssymb,
aps,prl]{revtex4-1}

\usepackage{graphicx}
\usepackage{dcolumn}
\usepackage{bm}
\usepackage{hyperref}
\usepackage{xcolor}
\usepackage{ulem}
\usepackage[mathlines]{lineno}
\usepackage[none]{hyphenat} 

\newcommand\delete{\bgroup\markoverwith{\textcolor{red}{\rule[0.5ex]{2pt}{0.4pt}}}\ULon}

\def\replaced{\@ifnextchar[{\xreplaced}{\yreplaced}}

\long\def\xreplaced[#1]#2#3{%
\ifdraft
\global\advance\refchangenumber by 1
\ifnumlines\xdef\doit{\noexpand\linelabel{\the\refchangenumber}}\doit\else%
\xdef\doit{\noexpand\label{\the\refchangenumber}}\doit\fi%
{\color{trackcolor}(Replaced: \sout{#2}}
{\color{black}replaced with:} {\color{trackcolor} #3)}%
\expandafter\gdef\csname
changenum\the\refchangenumber\endcsname{Replaced: [#1]
{\color{trackcolor}\sout{#2}} {\color{black} replaced with:}
{\color{trackcolor}#3}, }\else#3\fi}

\long\def\yreplaced#1#2{%
\ifdraft
\global\advance\refchangenumber by 1
\ifnumlines\xdef\doit{\noexpand\linelabel{\the\refchangenumber}}\doit\else%
\xdef\doit{\noexpand\label{\the\refchangenumber}}\doit\fi%
{\color{trackcolor}(Replaced: \sout{#1}}
{\color{black}replaced with:} {\color{trackcolor} #2)}%
\expandafter\gdef\csname changenum\the\refchangenumber\endcsname{Replaced:
{\color{trackcolor}\sout{#1}} {\color{black} replaced with:}
{\color{trackcolor}#2}, }\else#2\fi}

\begin{document}

\preprint{Fan \textit{et al.}}

\title{Directional radiation of sound waves by a subwavelength source}

\author{Xu-Dong Fan}
\author{Yi-Fan Zhu}
\author{Bin Liang}
\email{liangbin@nju.edu.cn}
\author{Jian-chun Cheng}
\email{jccheng@nju.edu.cn}
\affiliation{Collaborative Innovation Center of Advanced Microstructures and Key Laboratory of Modern Acoustics, MOE, Institute of Acoustics, Department of Physics, Nanjing University, Nanjing 210093, P. R. China}
\author{Likun Zhang}
\email{zhang@olemiss.edu}
\affiliation{National Center for Physical Acoustics and Department of Physics and Astronomy, University of Mississippi, University, Mississippi 38677, USA}

\date{\today}

\begin{abstract}
We propose and experimentally achieve a directional dipole field radiated by an omnidirectional monopole source enclosed in a subwavelength structure of acoustically hybrid resonances. The whole structure has its every dimension at an order smaller than the sound wavelength. The significance is that the radiation efficiency is up to 2.3 of the radiation by traditional dipole consisting of two out-of-phase monopoles in the same space. This study eventually takes an essential step towards solving the long-existing barrier of inefficient radiation of directional sound waves in low frequencies, and consequently inspires the ultimate radiation of arbitrary multipoles or even a highly directional beam by a monopole in limited spaces.
\end{abstract}

\pacs{XX}
\maketitle


Directional radiation of sound waves is of fundamental interest to the field of acoustics, with obvious potential to revolutionize various applications, such as focused sound waves for imaging, directional sound beams for underwater communication, among others. Yet in practice, from a classical perspective, any sound source with dimensions much smaller than the wavelength will act as a monopole and always radiate sound energy equally in all directions while the multipole radiation is inefficient relevant to the monopole\cite{ref:TA, ref:AJP-R1999}. To compensate the source size limitation, directional radiated fields at low frequency have to be produced via arrays of sources\cite{ref:PRA-Li2015} or sources that must move by long distance to synthesize a large-scale aperture in order for a source dimension comparable to the wavelength\cite{ref:JASA-AJ2016,ref:NDT-Azar2000,ref:IEEE-Wang1992,ref:IEEE-Ebbini1989,ref:JNE-Clay1999}. This poses a fundamental barrier on radiation of directional sound waves at low frequencies by a finite-size source, therefore preventing the effective downscaling of acoustic devices. We pursue radiation of low-frequency directional sound waves by a subwavelength source.

Many mechanisms have already been developed for controlling low-frequency sound radiation, which usually depend on coiling up space structures that force the sound to travel through a zigzag path to delay its propagation phase\cite{ref:APL-Zhao2013,ref:PRA-Li2014,ref:PRL-Liang2012,ref:NC-Zhu2016}, or some hybrid resonant structures that use arrays of resonators to accumulate a sufficient phase lag\cite{ref:JNE-Clay1999,ref:PRA-Li2015}. The existing metastructures based on these mechanisms, despite their vanishing thickness, still need to have a large transverse dimension in terms of wavelength to form a directional beam. On the other hand, metamaterials with anisotropic parameters designed by coordinate transformation technique can rotate the omni-directivity of a sound field but is still not able to reshape the directivity in subwavelength dimension\cite{ref:APL-Jiang2016}. Instead, we seek to develop a structure enclosing a monopole that as a whole has a subwavelength dimension yet can radiate directional sound waves at low frequency.

This letter takes the first step for breaking through the barrier to enable a high efficient production of a dipolar field from a deep-subwavelength structure wrapping around a simple point source. To elucidate our mechanism underlying such unusual phenomenon, we begin with a revisit of the sound field radiated from a finite-size source $q$ of a volume $V$, 
$\Phi (\mathbf{r})=\int_V {q({{\mathbf{r}}_1})G(\mathbf{r},{{\mathbf{r}}_1}){{d}^3}{{\mathbf{r}}_1}}$, 
where $G(\mathbf{r},{{\mathbf{r}}_1}){=}{{{\operatorname{e}}^{ik|\mathbf{r}-{{\mathbf{r}}_1}|}}}/{4\pi |\mathbf{r}-{{\mathbf{r}}_1}|}\;$is the Green's function of free space with $k$ being the wave number\cite{ref:TA}. By taking the multipole expansion around the source center $\mathbf{r} = \mathbf{r}_0$, 
\begin{equation}
\Phi (\mathbf{r})
=\int_{V}{q({{\mathbf{r}}_1})
G(\mathbf{r},{{\mathbf{r}}_0})
[1 + p_d({{\mathbf{r}}_1}) + ...]
{{d}^3}{{\mathbf{r}}_1}},
\end{equation}
where the first term is the monopole term, while the second term represents the dipole radiation with the factor $p_d$ in the far field form as
\begin{equation}
\label{eq:4}
p_d({{\mathbf{r}}_1})  = ({{\mathbf{r}}_1}-{{\mathbf{r}}_0})\cdot \nabla G(\mathbf{r},{{\mathbf{r}}_0}) / G(\mathbf{r},{{\mathbf{r}}_0})  \approx -ik | {\mathbf{r}}_1-{{\mathbf{r}}_0}| \cos\theta
\end{equation}
For a low-frequency airborne sound radiated in free space, one has $k|{\mathbf{r}}_1-{{\mathbf{r}}_0}|\ll1$ while radiated from a source of a small volume $V$ with $
V^{1/3}\ll\lambda$ ($\lambda$ is the wavelength), and hence the dipole part would be trivial as compared to the monopole part, in accordance with the conventional notion that any source with subwavelength dimension always behaves like a monopole with omni-directional radiation. A directional radiation of low-frequency sound, as a consequence, has to depend on a sufficiently large $| {\mathbf{r}}_1-{{\mathbf{r}}_0} |$ in terms of wavelength, which is realized by arranging small sources into an array with dimension comparable with wavelength\cite{ref:PRA-Li2015} or by synthesizing a large-scale aperture with moving sources, as done in traditional methods.\cite{ref:JASA-AJ2016,ref:NDT-Azar2000,ref:IEEE-Wang1992,ref:IEEE-Ebbini1989,ref:JNE-Clay1999} 

Here we achieve the radiation of a dipole field from a monopole source enclosed by a proposed structure with deep-subwavelength dimension, based on an essentially different mechanism that manipulates the effective wave number $k={k_{\mbox{eff}}}$ instead of enlarging the dimension of source for achieving a large $| {\mathbf{r}}_1-{{\mathbf{r}}_0} |$. We substantially expand the effective wave number $k_{\mbox{eff}}$ to make $k| {\mathbf{r}}_1-{{\mathbf{r}}_0} |\sim 1$, thereby satisfying the basic requirement that the wavelength and the source dimension must be comparable. We demonstrate this mechanism both numerically and experimentally by converting a monopole into a dipole with near-unity efficiency with a deep-subwavelength structure.

\begin{figure}
\centering
\includegraphics[width=8.3cm]{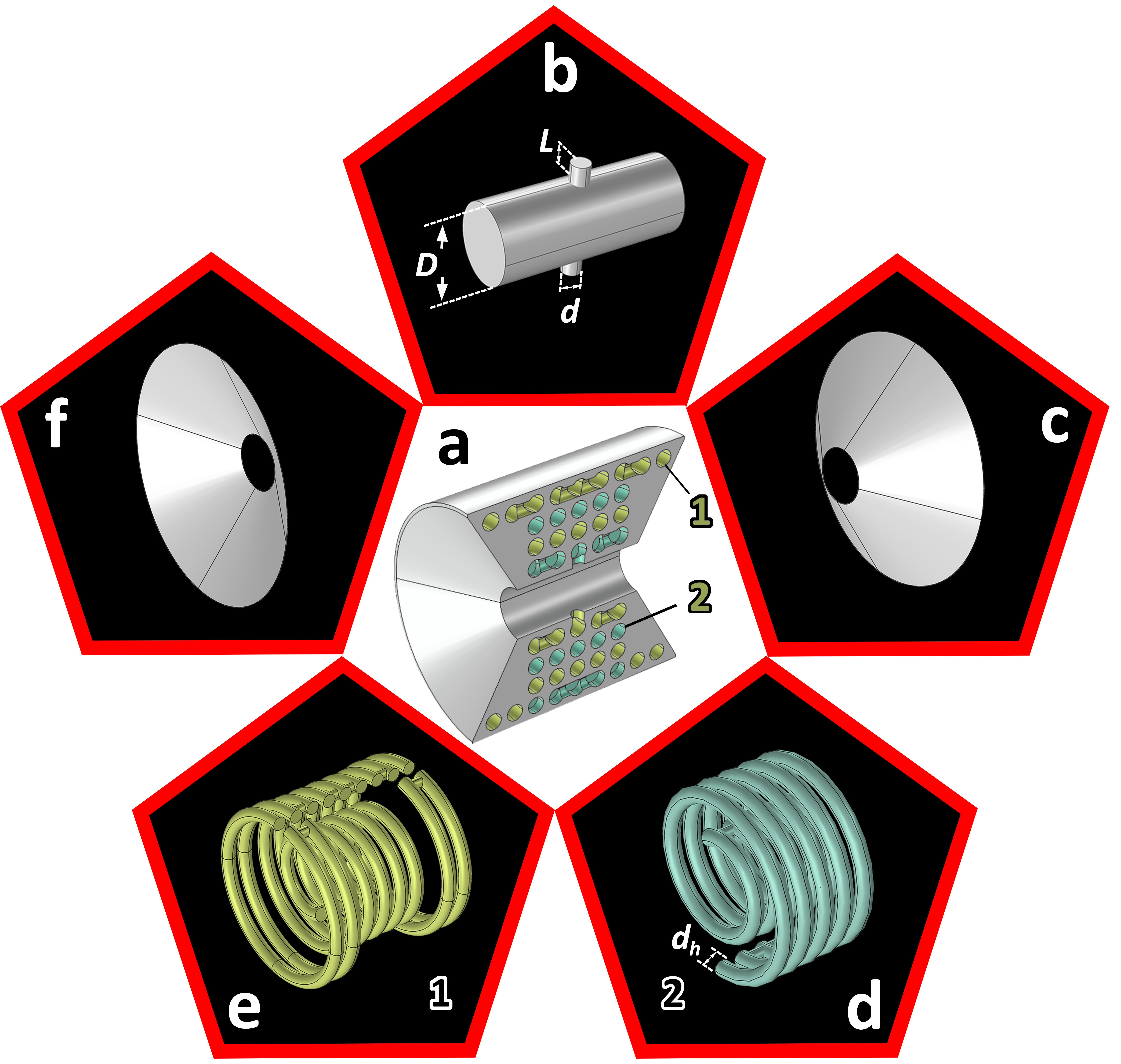}
\caption{\label{fig:1} Diagram of the artificial structure. The structure in (a) is composed of the several parts in (b-f).}
\end{figure}

\section*{Model}
The artificial structure to implement our distinctive idea, shown in \textbf{Fig.\ref{fig:1}(a)}, is formed by coupling a straight cylindrical tube of constant cross-sectional area $S$ shown in \textbf{Fig.\ref{fig:1}(b)} with two facing spiral tubes shown in \textbf{Fig.\ref{fig:1}(d)} and  \textbf{(e)}. The two facing tubes, spirally wrapped around the straight tube with their openings being attached at two opposite openings in the middle of that tube, maximally compress the dimension of the resulting device. And the two parts shown in \textbf{Fig.\ref{fig:1}(c)} and  \textbf{(f)} are connected to the straight cylindrical tube on the both ends to optimize the radiation by improving the impedance matching. While guaranteeing the deep subwavelength scale of the overall size of the whole structure, the combination of straight and spiral tubes supports hybrid internal resonances that substantially slow down the transmission wave and change the effective wave number $k={k_{\mbox{eff}}}$ to make $k| {\mathbf{r}}_1-{{\mathbf{r}}_0} |\sim 1$, as will be  proven in what follows. 

Consider the sound field radiated from a simple monopole source located at one end of the straight tube shown in \textbf{Fig.\ref{fig:1}(b)}. With a cross-sectional dimension $D\ (S={\pi {{D}^{2}}}/{4}\;)$ much smaller than the driving wavelength, this tube can be acoustically treated as a cylindrical waveguide allowing only the lowest $0^{th}$ mode to propagate. Without considering the near field effect caused by the approximation of a point source to a plane wave in the subwavelength waveguide, the sound pressure and volume velocity at the left end of the tube (i.e., the location of the point source) give:
${{p}_{0}}={{A}_{0}}\exp (i{{k}_{0}}x)+{{B}_{0}}\exp (-i{{k}_{0}}x)$, and
${{U}_{0}}=\frac{S}{\rho {{c}_{0}}}[{{A}_{0}}\exp (i{{k}_{0}}x)-{{B}_{0}}\exp (-i{{k}_{0}}x)]$, 
where ${{k}_{0}}$ is wave number in the air, ${{A}_{0}}$ and ${{B}_{0}}$ are the complex amplitudes of incident wave and reflected wave respectively. 
Correspondingly, the sound pressure and volume velocity at the other end of the tube gives:
${{p}_{2}}={{A}_{2}}\exp (i{{k}_{0}}x)$,  and 
${{U}_{2}}=\frac{S}{\rho {{c}_{0}}}{{A}_{2}}\exp (i{{k}_{0}}x)$.
At the junction, the continuity of pressure and velocity requires:
${{A}_{0}}+{{B}_{0}}={{A}_{2}}={{p}_{1}}$, and 
$\frac{S}{\rho {{c}_{0}}}({{A}_{0}}-{{B}_{0}})={{U}_{1}}+\frac{S}{\rho {{c}_{0}}}{{A}_{2}}$, 
where ${{p}_{1}}$ and ${{U}_{1}}$ are the sound pressure and volume velocity at the opening of the branch, respectively. It follows that the reflected coefficient and transmitted coefficient are
\begin{equation} 
{{r}_{p}}\equiv \frac{{{B}_{0}}}{{{A}_{0}}}=-\frac{{{Z}_{0}}}{{{Z}_{0}}+2{{Z}_{b}}}, ~~
{{t}_{p}}\equiv \frac{{{A}_{2}}}{{{A}_{0}}}=-\frac{2{{Z}_{b}}}{{{Z}_{0}}+2{{Z}_{b}}} \label{eq:17}
\end{equation}
where ${{Z}_{0}}={\rho {{c}_{0}}}/{S}$ is the acoustic impedance of the central tube, and ${{Z}_{b}}={{{Z}_{1}}{{Z}_{2}}}/{({{Z}_{1}}+{{Z}_{2}})}$ is the combined acoustic impedance of the two facing spiral tubes. $\rho $ and ${{c}_{0}}$ is the medium density and sound speed respectively, and ${{Z}_{1}}$, ${{Z}_{2}}$ are the acoustic impedances of the two branches, which are obtained by the impedance transfer equation to be:
\begin{subequations}
\begin{eqnarray} 
&&{{Z}_{1(2)}}=\frac{{{\rho }_{0}}{{c}_{0}}}{{{S}_{0}}}\frac{Z_{1(2)}^{*}-i\frac{{{\rho }_{0}}{{c}_{0}}}{{{S}_{0}}}\tan ({{k}_{0}}L)}{\frac{{{\rho }_{0}}{{c}_{0}}}{{{S}_{0}}}-iZ_{1(2)}^{*}\tan ({{k}_{0}}L)} \label{eq:18}\\
&& 
Z_{1(2)}^{*}=i\frac{{{\rho }_{0}}{{c}_{0}}}{{{S}_{h}}}\cot ({{k}_{0}}{{l}_{1(2)}})\label{eq:19}
\end{eqnarray}
\end{subequations}
where ${{S}_{0}}={\pi {{d}^{2}}}/{4}\;$ and ${{S}_{h}}={\pi {{d}_{h}^{2}}}/{4}\;$ are the cross-sectional areas of the neck shown in \textbf{Fig.\ref{fig:1}(b)} and the spiral tubes shown in \textbf{Fig.\ref{fig:1}(d)} and \textbf{(e)}, and $L$ and ${{l}_{1(2)}}$ are the lengths of the neck and the spiral tubes respectively. From the analytical formulae derived above, it can be anticipated that at the right end of the straight tube, the wave field would be of nearly equal strength and opposite phase as compared to the source located at the other end of that tube, if the structural parameters of straight and spiral tubes could be adjusted appropriately. This corresponds to a dramatic increase of the effective wave vector that ensures $k|{{r}_{1}}-{{r}_{0}}|\sim 1$ despite the deep subwavelength scale of the whole structure. The result is that, as can be predicted from Eqs.~(\ref{eq:4}), the original monopole term vanishes whereas the dipole term dominates, enabling a directional radiation pattern in far field that would otherwise be totally omni-directional under such circumstances. 

\section*{Simulations and measurements}
Now we simulate the directional radiation of low-frequency sound with the commercial COMSOL MULTIPHYSICS software based on the finite element method. The density and sound speed of artificial structure are set as $\rho =1250$~kg/m$^3$ and $c=2700$~m/s, respectively, and the background material applied are air. The standard parameters used for air under an ambient pressure of 1 atm at $20^\circ$~C are mass density $\rho_0 =1.21$~kg/m$^3$ and sound speed $c_0=343$~m/s. Perfectly matched layers (PMLs) are imposed on the outer boundaries of simulated domain to eliminate the interference from reflected wave.

\begin{figure}
\centering
\includegraphics[width=8.3cm]{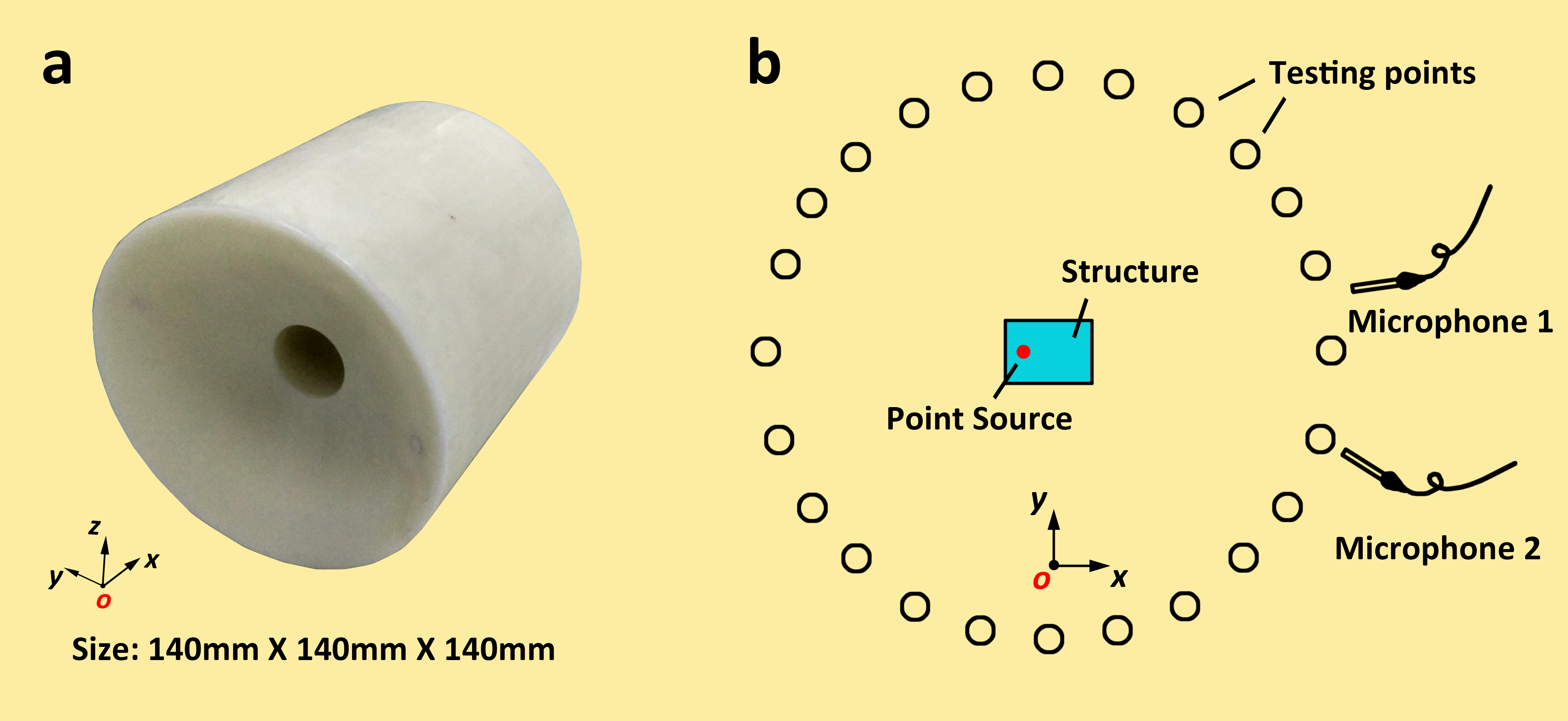}
\caption{\label{fig:2} Experiment setup. (a)Sample photo (b)Diagram of measurement environment.}
\end{figure}

Experiments to verify and demonstrate our conversion of monopole source into a dipole are conducted in anechoic chamber. \textbf{Figure \ref{fig:2}(a)} shows the sample of the artificial structure and the experiment setup is shown in \textbf{Fig.\ref{fig:2}(b)}. The solid sample is fabricated with polylactic acid (PLA) plastic via integrated 3D printing technology (Stratasys Dimension Elite, 0.08 mm in precision) to meet the requirement of the theoretical design. The material of sample is treated acoustically rigid due to the huge impedance mismatch between the solid and air. A loudspeaker (18 mm in length and 10mm in width) emitting a monochromatic wave as the acoustic source is placed inside the sample. For each measurement, two 1/4-in microphones (B\&K type 4961) are placed at designated positions to sense the local pressure: one is mounted at a fixed position to detect the pressure as signal 1 and the other is moveable to scan the pressure field as signal 2. By using the software PULSE LABSHOP, the cross spectrum of the two signals gathered by the two microphones is obtained, for which signal 1 works as a reference and signal 2 as an input signal. The pressure field is retrieved by analyzing the cross spectrum and recording the magnitude and phase at different spatial positions within the measured region.

\begin{figure}
\centering
\includegraphics[width=8.3cm]{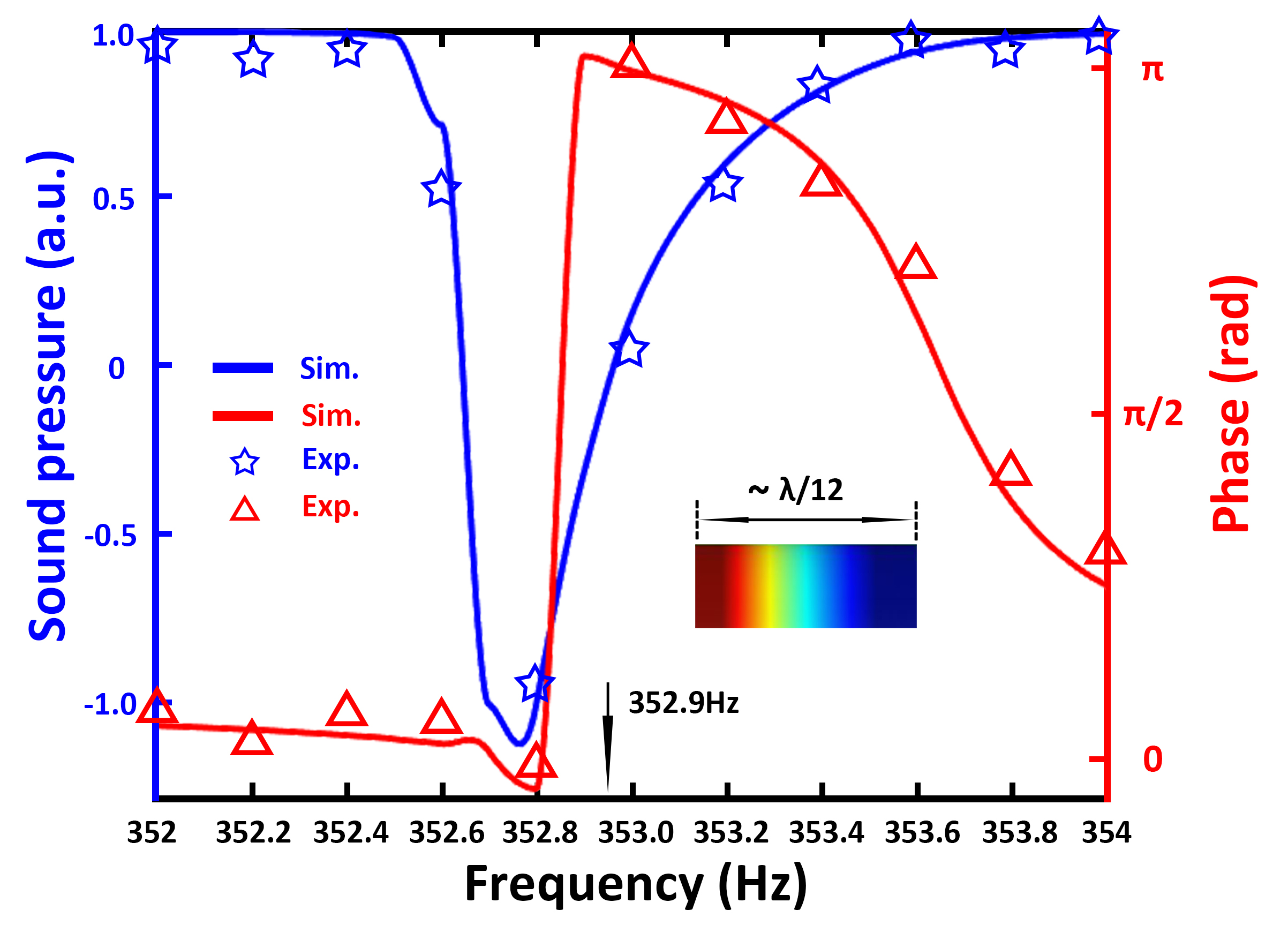}
\caption{\label{fig:4}Results of the amplitude difference and phase difference between the two openings of the artificial structure. Blue and red solid lines are simulated results of amplitude difference and phase difference versus frequency with star and triangle mark being the corresponding measured data. Black arrow points at around 352.9 Hz. (Insert: Sound pressure distribution inside the cylindrical tube at the resonant frequency)}
\end{figure}

\textbf{Figure \ref{fig:4}} shows the simulated results and measured data of the amplitude and phase differences on the two openings of the artificial structure with the sound pressure distribution inside the cylindrical tube at the resonant frequency being inserted. From the result, we can obtain that the amplitude difference is zero and the phase difference is $\pi $ at around 352.9Hz, which is the frequency to realize a dipole-like directivity as implied by the black arrow. And the corresponding ${L_{\mbox{structure}}/{\lambda}}=0.14$ (i.e., $kL_{\mbox{structure}}=0.9$) and ${L_{\mbox{tube}}/{\lambda}}=0.08$ (i.e., $kL_{\mbox{tube}}=0.5$). Good agreements between the measurement data, the simulation results and the theoretical prediction prove that our artificial structure can convert a monopole source to radiate sound waves like a dipole source in subwavelength dimension.

\begin{figure}
\centering·
\includegraphics[width=8.3cm]{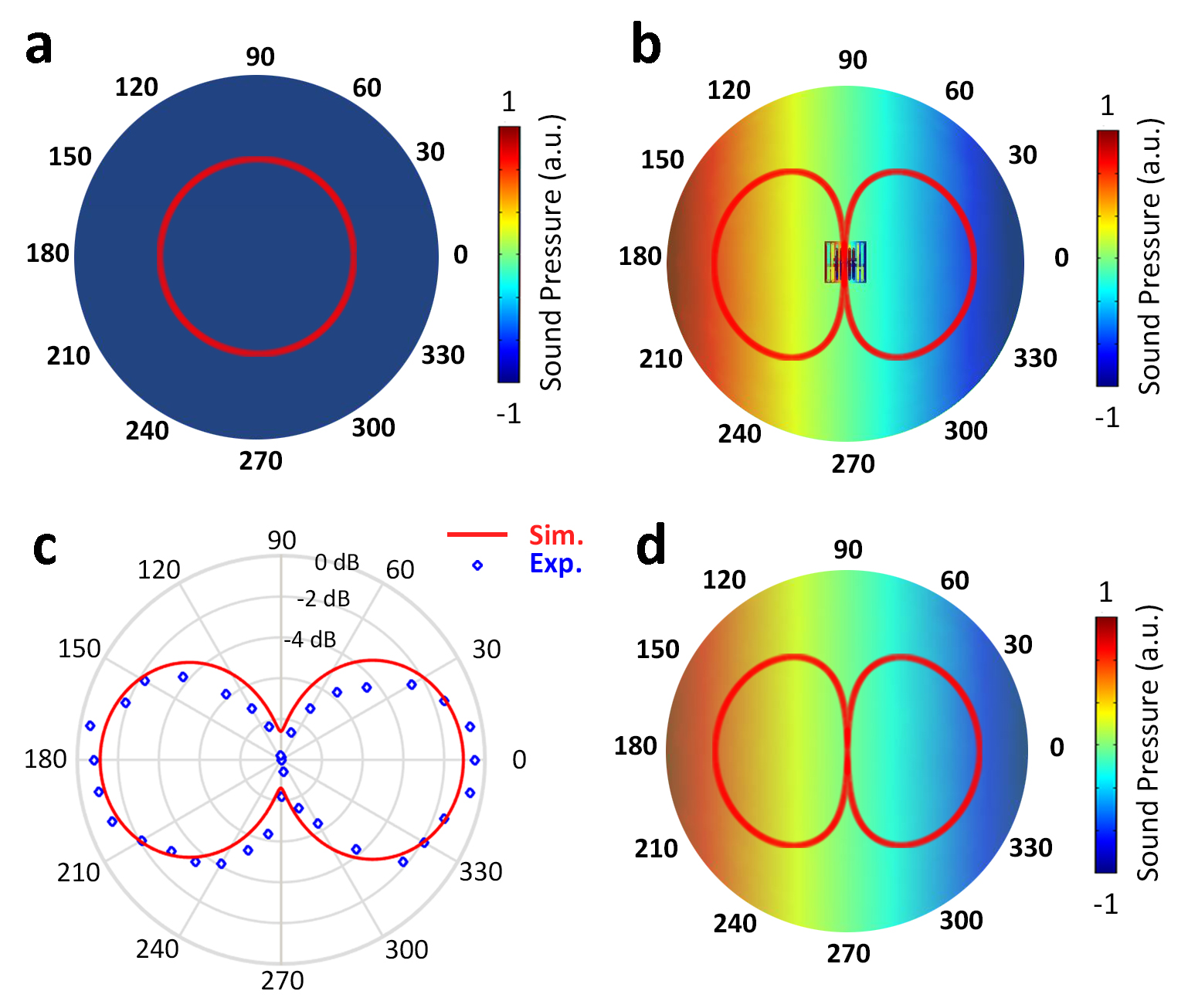}
\caption{\label{fig:3} Sound pressure fields and directivity patterns of (a) a monopole source without structure, (b) a monopole source with structure. (c) Simulated and measured directivity patterns of the artificial dipole. (d) Sound pressure fields and directivity patterns of a dipole source. The working frequency is around 352.9Hz.}
\end{figure}

\textbf{Figures \ref{fig:3}(a)} and \textbf{\ref{fig:3}(b)} display the simulated spatial distribution of pressure fields for sound radiated from low-frequency monopole sources with and without the designed artificial structure respectively. \textbf{Figure \ref{fig:3}(a)} shows that in the absence of any artificial structure, the monopole source radiates sound equally in all directions as expected. In the presence of our designed structure, the original radiation field produced by the monopole source is converted into a dipole with near-unity efficiency, as shown in \textbf{Fig.\ref{fig:3}(b)}, revealing the effectiveness of our scheme of generating directional radiation for low-frequency sound.

\textbf{Figure \ref{fig:3}(c)} shows the simulated results and the measured data of the directivity. Measured results agree well with the simulated ones, with both proving that our artificial structure can convert a monopole source to radiate sound waves like a dipole source in subwavelength dimension.

For a better comparison, we also plot the radiation of a simple dipole in free space in \textbf{Fig.\ref{fig:3}(d)}. 
The artificial dipole realized by the combination of a single monopole source and the subwavelength structure physically avoids the destructive interference of the two sources with opposite phases in a subwavelength interval. The radiation efficiency is 2.3 of the traditional dipole consisting of two out-of-phase monopole sources separated at a distance equal to the dimension of the structure ${L_{\mbox{structure}}}=0.14{\lambda}$. This amplification is beneficial from our structure that supports hybrid resonances in the tubes.

\section*{Discussion}
We have demonstrated the conversion of acoustic monopole source to dipole in a subwavelength source space via numerical simulations and experimental measurements. The whole structure has its every dimension at an order smaller than the sound wavelength, however, the radiation efficiency is up to 2.3 of the direct radiation of a dipole consisting of two out-of-phase monopoles seperated at a subwavelength distance equal to the dimension of our structure. 

The design is beneficial from the coupling resonances of sound waves in our structure. While it is expected from the known properties of acoustic resonances to provide phase shift in limited spaces \cite{ref:TA}, our structure has used a configuration of spiral tubes that enable the resonances to be effectively excited in a way that the coupling of sound energy in the structure leads to the desired phase delay for the efficient radiation of directional sound waves from an omnidirectional monopole enclosed by our structure that has a subwavelengh dimension as a whole.

The scheme proposed here only works within a narrow frequency band due to the mechanism of resonances, but the narrow-band sound source is also an important category of acoustic speakers/transducers that traditionally was associated with the resonances of the crystal of the transducers, though it is of interest to have a directional radiation of broad-band sources by a subwavelength source that we pursue here.  The results inspire the further study of converting monopole to arbitrary multipoles or even to a desired directional beam in subwavelength dimension with much higher radiation efficiency. Besides, the directional radiation of low frequency sound is also important to help understand the phenomena in nature such as for the dolphin to use directional sound for long-range communication even with its subwavelength vocal source. Our scheme takes the advantages of simpleness and compactness compared with existing traditional methods, which may open an avenue to solve the long-existing problem of omnidirectional radiation in low frequency in subwavelength dimension.

\begin{acknowledgments}
This work was supported by a start-up fund of the University of Mississippi, the National Natural Science Foundation of China (Grant Nos. 11634006 and 81127901), and A Project Funded by the Priority Academic Program Development of Jiangsu Higher Education Institutions.
\end{acknowledgments}

\bibliography{manuscript}

\begin{thebibliography}{13}%
\makeatletter
\providecommand \@ifxundefined [1]{%
 \@ifx{#1\undefined}
}%
\providecommand \@ifnum [1]{%
 \ifnum #1\expandafter \@firstoftwo
 \else \expandafter \@secondoftwo
 \fi
}%
\providecommand \@ifx [1]{%
 \ifx #1\expandafter \@firstoftwo
 \else \expandafter \@secondoftwo
 \fi
}%
\providecommand \natexlab [1]{#1}%
\providecommand \enquote  [1]{``#1''}%
\providecommand \bibnamefont  [1]{#1}%
\providecommand \bibfnamefont [1]{#1}%
\providecommand \citenamefont [1]{#1}%
\providecommand \href@noop [0]{\@secondoftwo}%
\providecommand \href [0]{\begingroup \@sanitize@url \@href}%
\providecommand \@href[1]{\@@startlink{#1}\@@href}%
\providecommand \@@href[1]{\endgroup#1\@@endlink}%
\providecommand \@sanitize@url [0]{\catcode `\\12\catcode `\$12\catcode
  `\&12\catcode `\#12\catcode `\^12\catcode `\_12\catcode `\%12\relax}%
\providecommand \@@startlink[1]{}%
\providecommand \@@endlink[0]{}%
\providecommand \url  [0]{\begingroup\@sanitize@url \@url }%
\providecommand \@url [1]{\endgroup\@href {#1}{\urlprefix }}%
\providecommand \urlprefix  [0]{URL }%
\providecommand \Eprint [0]{\href }%
\providecommand \doibase [0]{http://dx.doi.org/}%
\providecommand \selectlanguage [0]{\@gobble}%
\providecommand \bibinfo  [0]{\@secondoftwo}%
\providecommand \bibfield  [0]{\@secondoftwo}%
\providecommand \translation [1]{[#1]}%
\providecommand \BibitemOpen [0]{}%
\providecommand \bibitemStop [0]{}%
\providecommand \bibitemNoStop [0]{.\EOS\space}%
\providecommand \EOS [0]{\spacefactor3000\relax}%
\providecommand \BibitemShut  [1]{\csname bibitem#1\endcsname}%
\let\auto@bib@innerbib\@empty
\bibitem [{\citenamefont {C.}\ and\ \citenamefont {U.}(1968)}]{ref:TA}%
  \BibitemOpen
  \bibfield  {author} {\bibinfo {author} {\bibfnamefont {Morse P.~M.}\
  \bibnamefont {C.}}\ and\ \bibinfo {author} {\bibfnamefont {Ingard~K.}\
  \bibnamefont {U.}},\ }\bibfield  {title} {\enquote {\bibinfo {title}
  {Theoretical acoustics},}\ }\href@noop {} {\bibfield  {journal} {\bibinfo
  {journal} {Princeton university}\ } (\bibinfo {year} {1968})}\BibitemShut
  {NoStop}%
\bibitem [{\citenamefont {Russell}\ \emph {et~al.}(1999)\citenamefont
  {Russell}, \citenamefont {Titlow},\ and\ \citenamefont
  {Bemmen}}]{ref:AJP-R1999}%
  \BibitemOpen
  \bibfield  {author} {\bibinfo {author} {\bibfnamefont {Daniel~A.}\
  \bibnamefont {Russell}}, \bibinfo {author} {\bibfnamefont {Joseph~P.}\
  \bibnamefont {Titlow}}, \ and\ \bibinfo {author} {\bibfnamefont {Ya-Juan}\
  \bibnamefont {Bemmen}},\ }\bibfield  {title} {\enquote {\bibinfo {title}
  {Acoustic monopoles, dipoles, and quadrupoles: An experiment revisited},}\
  }\href {\doibase 10.1119/1.19349} {\bibfield  {journal} {\bibinfo  {journal}
  {American Journal of Physics}\ }\textbf {\bibinfo {volume} {67}},\ \bibinfo
  {pages} {660} (\bibinfo {year} {1999})}\BibitemShut {NoStop}%
\bibitem [{\citenamefont {Li}\ \emph {et~al.}(2015)\citenamefont {Li},
  \citenamefont {Jiang}, \citenamefont {Liang}, \citenamefont {Jianchun},\ and\
  \citenamefont {Zhang}}]{ref:PRA-Li2015}%
  \BibitemOpen
  \bibfield  {author} {\bibinfo {author} {\bibfnamefont {Yong}\ \bibnamefont
  {Li}}, \bibinfo {author} {\bibfnamefont {Xue}\ \bibnamefont {Jiang}},
  \bibinfo {author} {\bibfnamefont {Bin}\ \bibnamefont {Liang}}, \bibinfo
  {author} {\bibfnamefont {Cheng}\ \bibnamefont {Jianchun}}, \ and\ \bibinfo
  {author} {\bibfnamefont {Likun}\ \bibnamefont {Zhang}},\ }\bibfield  {title}
  {\enquote {\bibinfo {title} {Metascreen-based acoustic passive phased
  array},}\ }\href {\doibase 10.1103/PhysRevApplied.4.024003} {\bibfield
  {journal} {\bibinfo  {journal} {Phys. Rev. Appl.}\ }\textbf {\bibinfo
  {volume} {4}},\ \bibinfo {pages} {024003} (\bibinfo {year}
  {2015})}\BibitemShut {NoStop}%
\bibitem [{\citenamefont {Berkhout}\ \emph {et~al.}(2016)\citenamefont
  {Berkhout}, \citenamefont {Vries},\ and\ \citenamefont
  {P.}}]{ref:JASA-AJ2016}%
  \BibitemOpen
  \bibfield  {author} {\bibinfo {author} {\bibfnamefont {A.~J.}\ \bibnamefont
  {Berkhout}}, \bibinfo {author} {\bibfnamefont {D.~de}\ \bibnamefont {Vries}},
  \ and\ \bibinfo {author} {\bibfnamefont {Vogel}\ \bibnamefont {P.}},\
  }\bibfield  {title} {\enquote {\bibinfo {title} {Acoustic control by wave
  field synthesis},}\ }\href {\doibase 10.1121/1.405852} {\bibfield  {journal}
  {\bibinfo  {journal} {The Journal of the Acoustical Society of America}\
  }\textbf {\bibinfo {volume} {93}},\ \bibinfo {pages} {2764} (\bibinfo {year}
  {2016})}\BibitemShut {NoStop}%
\bibitem [{\citenamefont {Azar}\ \emph {et~al.}(2000)\citenamefont {Azar},
  \citenamefont {Y.},\ and\ \citenamefont {Wooh}}]{ref:NDT-Azar2000}%
  \BibitemOpen
  \bibfield  {author} {\bibinfo {author} {\bibfnamefont {L.}~\bibnamefont
  {Azar}}, \bibinfo {author} {\bibfnamefont {SHi}\ \bibnamefont {Y.}}, \ and\
  \bibinfo {author} {\bibfnamefont {S.~C.}\ \bibnamefont {Wooh}},\ }\bibfield
  {title} {\enquote {\bibinfo {title} {Beam focusing behavior of linear phased
  arrays},}\ }\href {\doibase 10.1016/S0963-8695(99)00043-2} {\bibfield
  {journal} {\bibinfo  {journal} {NDT and E International}\ }\textbf {\bibinfo
  {volume} {33}},\ \bibinfo {pages} {189} (\bibinfo {year} {2000})}\BibitemShut
  {NoStop}%
\bibitem [{\citenamefont {Wang}(1992)}]{ref:IEEE-Wang1992}%
  \BibitemOpen
  \bibfield  {author} {\bibinfo {author} {\bibfnamefont {H.S.C.}\ \bibnamefont
  {Wang}},\ }\bibfield  {title} {\enquote {\bibinfo {title} {Performance of
  phased-array antennas with mechanical errors},}\ }\href {\doibase
  10.1109/7.144579} {\bibfield  {journal} {\bibinfo  {journal} {IEEE
  Transactions on Aerospace and Electronic Systems}\ }\textbf {\bibinfo
  {volume} {28}},\ \bibinfo {pages} {535} (\bibinfo {year} {1992})}\BibitemShut
  {NoStop}%
\bibitem [{\citenamefont {S.}\ and\ \citenamefont
  {A.}(1989)}]{ref:IEEE-Ebbini1989}%
  \BibitemOpen
  \bibfield  {author} {\bibinfo {author} {\bibfnamefont {Ebbini~E.}\
  \bibnamefont {S.}}\ and\ \bibinfo {author} {\bibfnamefont {Cain~C.}\
  \bibnamefont {A.}},\ }\bibfield  {title} {\enquote {\bibinfo {title}
  {Multiple-focus ultrasound phased-array pattern synthesis: optimal
  driving-signal distributions for hyperthermia},}\ }\href {\doibase
  10.1109/58.31798} {\bibfield  {journal} {\bibinfo  {journal} {IEEE
  transactions on ultrasonics, ferroelectrics, and frequency control}\ }\textbf
  {\bibinfo {volume} {36}},\ \bibinfo {pages} {540} (\bibinfo {year}
  {1989})}\BibitemShut {NoStop}%
\bibitem [{\citenamefont {C.}\ \emph {et~al.}(1999)\citenamefont {C.},
  \citenamefont {C.}, \citenamefont {L.},\ and\ \citenamefont
  {Wang}}]{ref:JNE-Clay1999}%
  \BibitemOpen
  \bibfield  {author} {\bibinfo {author} {\bibfnamefont {Clay~A.}\ \bibnamefont
  {C.}}, \bibinfo {author} {\bibfnamefont {Wooh~S.}\ \bibnamefont {C.}},
  \bibinfo {author} {\bibfnamefont {Azar}\ \bibnamefont {L.}}, \ and\ \bibinfo
  {author} {\bibfnamefont {Jiyong}\ \bibnamefont {Wang}},\ }\bibfield  {title}
  {\enquote {\bibinfo {title} {Experimental study of phased array beam steering
  characteristics},}\ }\href {\doibase 10.1023/A:1022618321612} {\bibfield
  {journal} {\bibinfo  {journal} {Journal of Nondestructive Evaluation}\
  }\textbf {\bibinfo {volume} {18}},\ \bibinfo {pages} {59} (\bibinfo {year}
  {1999})}\BibitemShut {NoStop}%
\bibitem [{\citenamefont {Zhao}\ \emph {et~al.}(2013)\citenamefont {Zhao},
  \citenamefont {Li}, \citenamefont {Chen},\ and\ \citenamefont
  {Qiu}}]{ref:APL-Zhao2013}%
  \BibitemOpen
  \bibfield  {author} {\bibinfo {author} {\bibfnamefont {Jiajun}\ \bibnamefont
  {Zhao}}, \bibinfo {author} {\bibfnamefont {Baowen}\ \bibnamefont {Li}},
  \bibinfo {author} {\bibfnamefont {Zhining}\ \bibnamefont {Chen}}, \ and\
  \bibinfo {author} {\bibfnamefont {ChengWei}\ \bibnamefont {Qiu}},\ }\bibfield
   {title} {\enquote {\bibinfo {title} {Redirection of sound waves using
  acoustic metasurface},}\ }\href {\doibase 10.1063/1.4824758} {\bibfield
  {journal} {\bibinfo  {journal} {Appl. Phys. Lett.}\ }\textbf {\bibinfo
  {volume} {103}},\ \bibinfo {pages} {151604} (\bibinfo {year}
  {2013})}\BibitemShut {NoStop}%
\bibitem [{\citenamefont {Li}\ \emph {et~al.}(2014)\citenamefont {Li},
  \citenamefont {Jiang}, \citenamefont {Li}, \citenamefont {Liang},
  \citenamefont {Zou}, \citenamefont {Yin},\ and\ \citenamefont
  {Cheng}}]{ref:PRA-Li2014}%
  \BibitemOpen
  \bibfield  {author} {\bibinfo {author} {\bibfnamefont {Yong}\ \bibnamefont
  {Li}}, \bibinfo {author} {\bibfnamefont {Xue}\ \bibnamefont {Jiang}},
  \bibinfo {author} {\bibfnamefont {Ruiqi}\ \bibnamefont {Li}}, \bibinfo
  {author} {\bibfnamefont {Bin}\ \bibnamefont {Liang}}, \bibinfo {author}
  {\bibfnamefont {Xinye}\ \bibnamefont {Zou}}, \bibinfo {author} {\bibfnamefont
  {Leilei}\ \bibnamefont {Yin}}, \ and\ \bibinfo {author} {\bibfnamefont
  {Jianchun}\ \bibnamefont {Cheng}},\ }\bibfield  {title} {\enquote {\bibinfo
  {title} {Experimental realization of full control of reflected waves with
  subwavelength acoustic metasurfaces},}\ }\href {\doibase
  10.1103/PhysRevApplied.2.064002} {\bibfield  {journal} {\bibinfo  {journal}
  {Phys. Rev. Appl.}\ }\textbf {\bibinfo {volume} {2}},\ \bibinfo {pages}
  {064002} (\bibinfo {year} {2014})}\BibitemShut {NoStop}%
\bibitem [{\citenamefont {Liang}\ and\ \citenamefont
  {Li}(2012)}]{ref:PRL-Liang2012}%
  \BibitemOpen
  \bibfield  {author} {\bibinfo {author} {\bibfnamefont {Zixian}\ \bibnamefont
  {Liang}}\ and\ \bibinfo {author} {\bibfnamefont {Jensen}\ \bibnamefont
  {Li}},\ }\bibfield  {title} {\enquote {\bibinfo {title} {Extreme acoustic
  metamaterial by coiling up space},}\ }\href {\doibase
  10.1103/PhysRevLett.108.114301} {\bibfield  {journal} {\bibinfo  {journal}
  {Phys. Rev. Lett.}\ }\textbf {\bibinfo {volume} {108}},\ \bibinfo {pages}
  {114301} (\bibinfo {year} {2012})}\BibitemShut {NoStop}%
\bibitem [{\citenamefont {Zhu}\ \emph {et~al.}(2016)\citenamefont {Zhu},
  \citenamefont {Li}, \citenamefont {Zhang}, \citenamefont {Zhu}, \citenamefont
  {Zhang}, \citenamefont {Tian},\ and\ \citenamefont {Liu}}]{ref:NC-Zhu2016}%
  \BibitemOpen
  \bibfield  {author} {\bibinfo {author} {\bibfnamefont {Xuefeng}\ \bibnamefont
  {Zhu}}, \bibinfo {author} {\bibfnamefont {Kun}\ \bibnamefont {Li}}, \bibinfo
  {author} {\bibfnamefont {Peng}\ \bibnamefont {Zhang}}, \bibinfo {author}
  {\bibfnamefont {Jie}\ \bibnamefont {Zhu}}, \bibinfo {author} {\bibfnamefont
  {Jintao}\ \bibnamefont {Zhang}}, \bibinfo {author} {\bibfnamefont {Chao}\
  \bibnamefont {Tian}}, \ and\ \bibinfo {author} {\bibfnamefont {Shengchun}\
  \bibnamefont {Liu}},\ }\bibfield  {title} {\enquote {\bibinfo {title}
  {Implementation of dispersion-free slow acoustic wave propagation and phase
  engineering with helical-structured metamaterials},}\ }\href {\doibase
  10.1038/ncomms11731} {\bibfield  {journal} {\bibinfo  {journal} {Nature
  Commun.}\ }\textbf {\bibinfo {volume} {7}},\ \bibinfo {pages} {11731}
  (\bibinfo {year} {2016})}\BibitemShut {NoStop}%
\bibitem [{\citenamefont {Jiang}\ \emph {et~al.}(2016)\citenamefont {Jiang},
  \citenamefont {Zhang}, \citenamefont {Liang}, \citenamefont {Zou},\ and\
  \citenamefont {Cheng}}]{ref:APL-Jiang2016}%
  \BibitemOpen
  \bibfield  {author} {\bibinfo {author} {\bibfnamefont {Xue}\ \bibnamefont
  {Jiang}}, \bibinfo {author} {\bibfnamefont {Likun}\ \bibnamefont {Zhang}},
  \bibinfo {author} {\bibfnamefont {Bin}\ \bibnamefont {Liang}}, \bibinfo
  {author} {\bibfnamefont {Xinye}\ \bibnamefont {Zou}}, \ and\ \bibinfo
  {author} {\bibfnamefont {Jianchun}\ \bibnamefont {Cheng}},\ }\bibfield
  {title} {\enquote {\bibinfo {title} {Radiation directivity rotation by
  acoustic metamaterials},}\ }\href {\doibase 10.1063/1.4930061} {\bibfield
  {journal} {\bibinfo  {journal} {Appl. Phys. Lett.}\ }\textbf {\bibinfo
  {volume} {107}},\ \bibinfo {pages} {093506} (\bibinfo {year}
  {2016})}\BibitemShut {NoStop}%
\end{thebibliography}%

\end{document}